# Forecast of the evolution of the contagious disease caused by novel coronavirus (2019-nCoV) in China

Javier Gamero [a], Juan A. Tamayo [b] and Juan A. Martínez-Román [a]

[a] Department of Applied Economics. University of Seville
[b] Department of Business Administration and Marketing. University of Seville
(corresponding author, e-mail address: jtamayo@us.es)

## Abstract

The outbreak of novel coronavirus (2019-nCoV) in China has caused a viral epidemic affecting tens of thousands of persons. Though the danger of this epidemic is evident, the statistical analysis of data offered in this paper indicates that the increase in new cases in China will stabilize in the coming days or weeks. Our forecast could serve to evaluate risks and control the evolution of this disease.

## 1. Introduction

Since December 30[th] 2019, when Dr. Li Wenliang and various colleagues alerted of the possible existence of an epidemic, there has been an ever increasing number of cases and deaths, including that of Li himself [1–3]. To avoid the spread of the disease, tens of millions of persons have been quarantined in the vicinity of Wuhan, and efforts have included such impressive feats as the construction of a thousand-bed hospital in just ten days. Despite these endeavors, the disease's advance and the level of risk detected by World Health Organization experts led the organization to declare novel coronavirus an "international emergency."

Though it is too soon to have an in-depth knowledge of the disease and its implications for public health, economies, etc., the data available allow for an initial forecast of the evolution of 2019-nCoV in China. This paper's objective is to provide a forecast on the evolution of confirmed cases in China. We hope that the information provided contributes to evaluating risks and controlling the epidemic's spread.

We will now proceed to the empirical analysis of known data on the number of cases in China; the statistic model shows that transmission will soon stabilize. Finally, we will draw some conclusions.

## 2. Empirical analysis

The techniques employed to study epidemics are diverse, including time series analysis [4–6], the comparison and application of various Grey Models [6–9], a univariate and stepwise multivariate linear regression [10], and the use of Markow chain prediction models [11,12] and simulation models [13,14]. In this paper, we analyze the temporal series of confirmed cases through a first order polynomial for the first difference of the series of daily accumulated confirmed cases expressed as a logarithm.

Calculations were carried out through a program developed in J language. J is an array programming language based on APL and developed by Kenneth E. Iverson.

### 2.1. Data

Data were obtained from daily reports of China's National Health Commission (NHC) as published in Wikipedia.

### 2.2. Study and visual representation of data

When predicting the evolution of a temporal process, it is useful to find a model generating past values in order to extrapolate said model to the future. To this effect, the model must sufficiently describe the past, be robust in its predictions, and fulfill the Occam's razor principle.

The temporal series we aim to predict is the number of accumulated confirmed cases (C) of virus 2019-nCoV in China. The temporal unit is a day and the numbers of cases are those provided by the NHC. A first approach to said series confirms that the logarithmic transformation allows us to describe the phenomenon in a functionally simpler manner and with the added advantage of expressing the relative or percentual evolution of number of cases in logarithmic form.

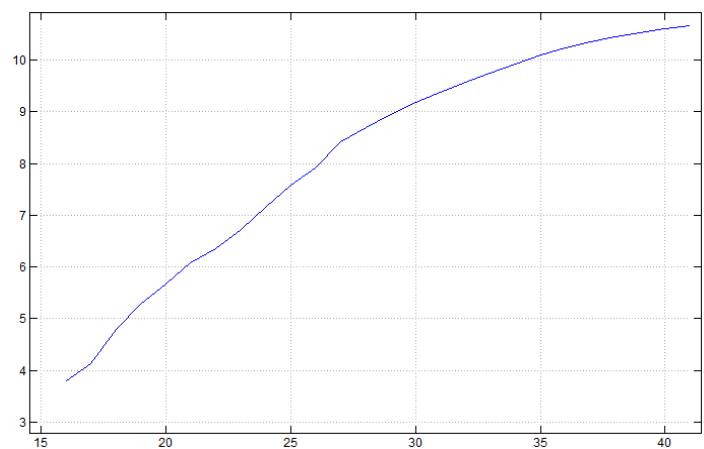

Figure 1. Logarithm of accumulated confirmed cases.

This series ln(C) appears at first glance to suggest a non-linear evolution close to the quadratic one. It should be pointed out that the quadratic evolution would only describe the phenomenon up to the peak, and following that point, would no longer be the generating model, given that C cannot decrease. It is a lifecycle model where the maximum is generated in a finite time and not in an infinite time like a classic logistic model.

This gives rise to two ideas: to try to directly adjust the quadratic model to this series, or to find the first difference in the series and adjust a linear model to said difference. In principle, a linear model would allow us to describe more simply the internal mechanism of the phenomenon's evolution. On the other hand, the quadratic adjustment shows considerably self-correlated residues that suggest that the series may not be adequately predictable through a simple model. Thus, we consider that the first analysis to attempt is the linear model on the first difference in the series.

The graph shows series dlC (=d(ln(C))) with a superimposed linear adjustment. In principle, it appears to be an adequate description of said difference. However, it is necessary to carry out some statistical verifications in order to validate the visual representation. We did two types of verifications:

• Verification of the linear adjustment: Analysis of the self-correlations of the residues and crossed validation of the degree of the adjustment polynomial.

• Robustness check in the predictions provided by the model: Stability in the predictions as days of observation increase and stability in the solutions as less importance is assigned to older data potentially less relevant to the future.

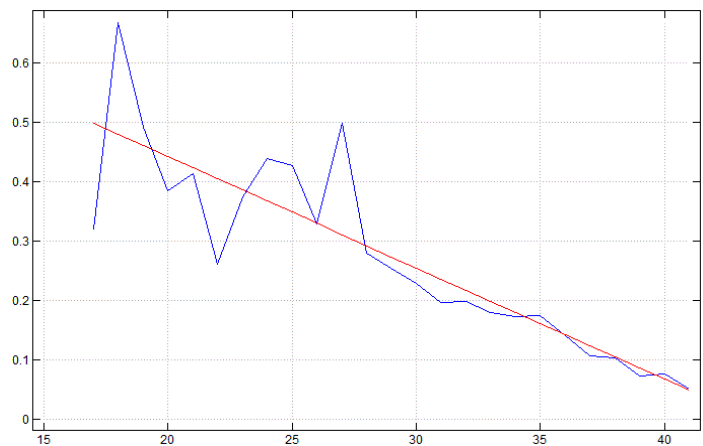

Figure 2. Difference in ln(C) and linear regression.

We calculated the first 10 self-correlations of the residues of the linear adjustment of d(ln(Cc)). Once standardized, no value significantly deviated from 0, such that there appears to be no underlying model more complex than the linear one to serve as generating model of past data. Obviously, the set of data is small (25 days), and this makes it more likely for the populational self-correlations to be null. As a reference, if residues in fact have null self-correlations, the first self- correlation observed could be in the interval (–0.467, 0.467), with a 95% confidence level. In any case, the Occam's razor principle indicates that we should not discard a simple model that could adequately describe the data.

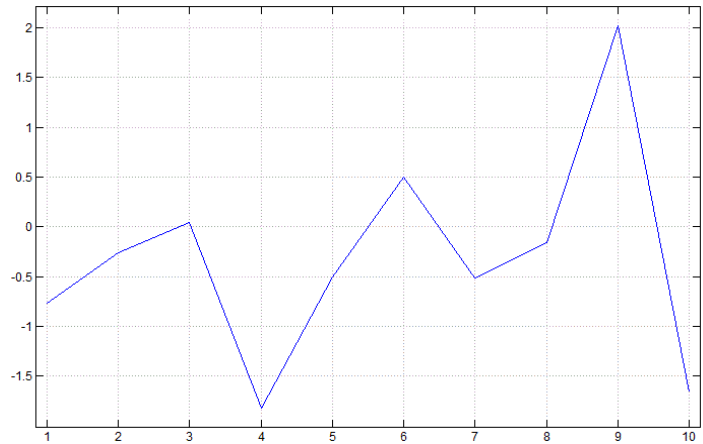

Figure 3. Self-correlations of dlCC expressed in z-values.

We have used the procedure of one-point cross-validation to check what degree of polynomial would be ideal for adjusting these data. The description is the following: for each polynomial degree, an adjustment was made by removing one of the observed data and noting the error in the estimation. This was carried out for all of the data individually and the quadratic error of all those errors was calculated. The values reflected in the Table are those quadratic errors for each degree from 0 to 5. The first-degree adjustment is the most logical choice.

| Degree | Value of the One-point cross-validation |
|--------|------------------------------------------|
| 0 | 0.0260 |
| 1 | 0.0071 |
| 2 | 0.0084 |
| 3 | 0.0098 |
| 4 | 0.0130 |
| 5 | 0.0218 |

Table 1. Results of the one-point cross-validation for different degrees of polynomial adjustment.

The linear adjustment obtained is:

$dlC(t) \cong 0.8166 - 0.01872 \cdot t$, t being the number of the day in year 2020 (February 1 = 32).

From this adjustment, we immediately obtain a prediction of t = 43.63 (February 12-13) for the moment of growth zero, in other words, the moment in which the quadratic process ln(C) reaches its maximum value, which in turn indicates when the number of accumulated confirmed cases C would stop growing. Obviously, starting at this point, the model would cease to work, given that number C by its very nature cannot decrease.

At the same time, it is possible to calculate the maximum value that would be reached in series C using the extrapolation of the model. In this case, the maximum of accumulated confirmed cases (obtained in t = 43.63) would be C = 45090.

We carried out robustness verifications for both predictions of the model (day and maximum value of C, in other words, the day on which virtually no confirmed cases of the disease will appear) in the two ways previously described.

The progressive stability of both predictions can be observed as the days pass, to the point where since approximately the beginning of February the predictions have not shown relevant variations.

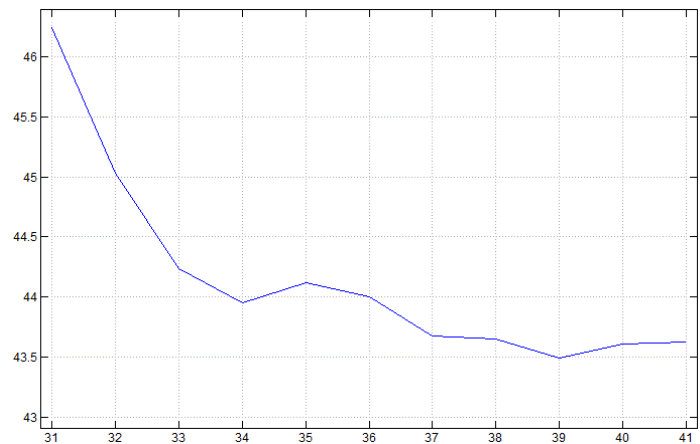

Figure 4. Predictions of the final day following changes in said prediction with data available on recent days.

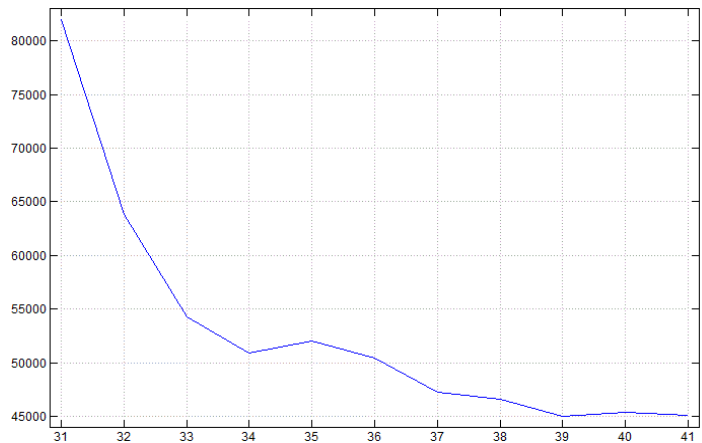

Figure 5. Predictions of the maximum of accumulated confirmed cases as they have changed with the data available on past days.

In any dynamic process, data closer in time to the moment to be predicted are potentially more relevant than other, more remote values. In our previously executed linear adjustment, all data were granted the same weight. It is worth asking whether the results would have been different if older data were given a smaller weight. To this end, we have made predictions with linear adjustments assigning decreasing weight to older data. We decreased the weightings exponentially by factor α per day passed, such that α = 1 indicates non-decrease (our linear model) and α = 0.7 implies an underweighting of 30% with each day that passes.

The graphs show the variations of the predictions of the day and the maximum value of C according to the degree of underweighting. Despite the fact that 0.7 is a severe underweighting and that with more underweighting, the equivalent sample becomes smaller, there would have been an insignificant variation in both predictions if we had applied a discount rate in our linear model.

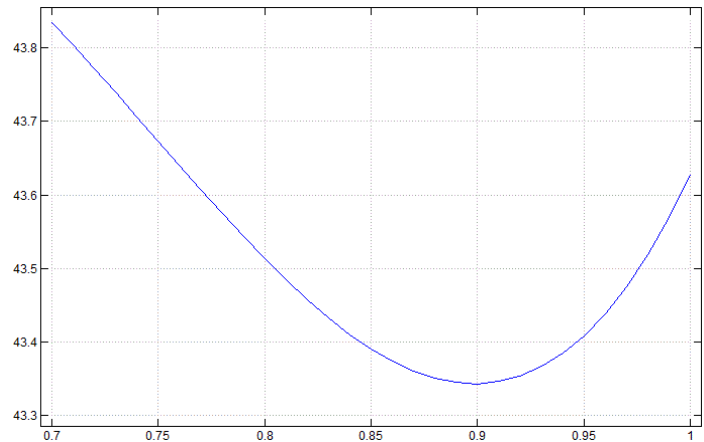

Figure 6. Prediction of the final day according to the underweighting of earlier days.

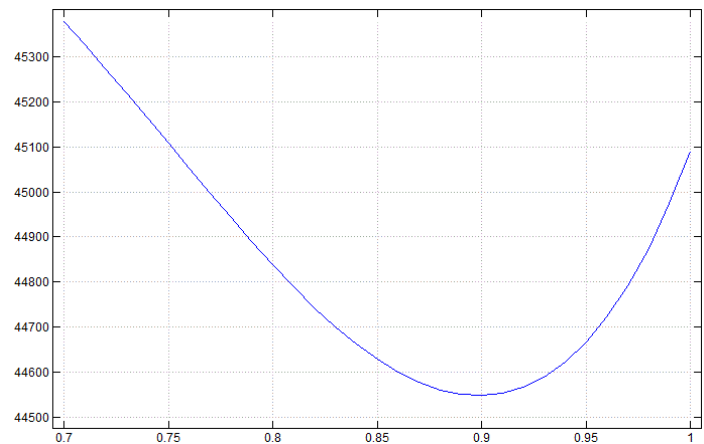

Figure 7. Prediction of maximum accumulated confirmed cases according to the underweighting of earlier days.

Both robustness verifications suggest that the predictions of our model behave well in the presence of possible changes in the data to be analyzed.

## 3. Conclusions

The model shows that the growth in confirmed cases will conclude in a matter of days or weeks, with a limited number of persons affected.

The alarming predictions made at the outset of this epidemic are fortunately not materializing. Thus, it appears that this is not a major epidemic in number of persons affected or extension in time, but rather an epidemic with a moderate impact.

Erroneous forecasts could imply a squandering of resources and could undermine citizens' confidence [15]. It is important to continue monitoring the evolution in number of persons affected in order to reevaluate the model and perceive over time the real scope of any deviation or change in the tendency of this epidemic.

## Bibliografía